\documentclass[letter]{aa}
\usepackage{txfonts}
\usepackage{graphicx}
\usepackage{longtable}
\usepackage{lscape}

\newcommand{\e}{et al.\ }
\newcommand{\pmra}{$\mu_{\alpha}$}
\newcommand{\pmdec}{$\mu_{\delta}$}
\newcommand{\gv}{$\gamma^2$~Vel}

\begin{document}

\title{Multiple kinematical populations in Vela OB2 from Gaia DR1 data
}

\date{Received date / Accepted date}

\author{F. Damiani\inst{1},
L. Prisinzano\inst{1},
R.~D. Jeffries\inst{2},
G.~G. Sacco\inst{3},
S. Randich\inst{3},
\and
G. Micela\inst{1}
}
\institute{INAF - Osservatorio Astronomico di Palermo G.S.Vaiana,
Piazza del Parlamento 1, I-90134 Palermo, Italy
\and
Astrophysics Group, Keele University, Keele, Staffordshire ST5 5BG,
United Kingdom
\and
INAF - Osservatorio Astrofisico di Arcetri, Largo E. Fermi
5, 50125, Firenze, Italy
}

\abstract{
Recent results using radial-velocity measurements from the Gaia-ESO Survey
have led to the discovery of
multiple kinematic populations across the Vela OB2 association. We
present here a proper-motion study of the same region.
}{Our aim is to test whether the radial-velocity populations have a
counterpart in proper-motion space, and if so, how the two sets of
kinematical data complement each other.
}{This work is based on parallaxes and proper motions from the TGAS catalog,
as part of Gaia DR1.
}{Two distinct proper-motion populations are found dispersed across
$\sim 5$ degrees (or $\sim 30$ pc at their likely distances).
Their detailed
correspondence
to the radial-velocity populations could not be tested, because of the
paucity of common objects. However, compelling indications are found that one
of the
new proper-motion populations consists mostly of members of the young
cluster NGC~2547, and the other is related to the \gv\ cluster.
Constraints on the age of the two populations, both of which
appear to be only 10-35~Myr old, and their possible
mutual interactions within the last 1.5~Myrs, are discussed.
}{}


\titlerunning{Kinematical populations in Vel OB2 from Gaia DR1 data}
\authorrunning{Damiani et al.}

\maketitle

\section{Introduction}
\label{intro}

The Vela OB2 association spans a wide sky region, inside the Gum nebula
(Pettersson 2008), and has been subject to an increasing number of studies
in recent years.  Its visually brightest member, the O8~III+Wolf-Rayet
binary star \gv,
was found to be surrounded by a cluster of lower-mass stars through X-ray
observations (Pozzo \e 2000).  The \gv\ cluster was then studied, among
others, by Jeffries \e (2009), and more recently by Jeffries \e (2014,
2017),
as part of the Gaia-ESO Survey (Randich \e 2013, Gilmore \e 2012). This
Survey provided precise radial-velocity data for hundreds of candidate
members of the \gv\ cluster, enabling Jeffries \e (2014) to find clear
evidence of a double kinematical population, separated by $\simeq 2$
km/s in radial velocity (termed Populations A and
B) among the young (somewhere between 10-20 Myr -- Jeffries et al.\ 2017),
pre-main-sequence low-mass stars in the cluster.
{
Population A was more spatially concentrated around $\gamma^2$ Vel and
had a significantly lower radial velocity dispersion than the more
uniformly located population B.
}
The sky region covered in this study had a size of approximately one
square degree, thus much smaller than the size of the entire Vela OB2
association, as found e.g.\ by the Hipparcos-based study of de Zeeuw \e
(1999).  Slightly later, still using data from the Gaia-ESO Survey, Sacco
\e (2015) found a similar duplicity in the kinematics of members of the
young cluster NGC~2547 (age $35 \pm3$ Myr, Jeffries and Oliveira 2005;
distance $361^{+19}_{-8}$~pc, Naylor and Jeffries 2006),
lying only $\sim$2 degrees south of \gv\ 
(distance $356^{+12}_{-11}$~pc, Jeffries \e 2009).
{
Two kinematic populations were found within 30 arcminutes of NGC~2547;
the dominant one was associated with the cluster, but the second sparser
group, separated in radial velocity by  $\simeq 6$ km/s, appears to be
similar in age and kinematics to group B of the younger \gv\ cluster.
}
Unfortunately, no additional
Gaia-ESO Survey observations exist towards Vela OB2, most of which remains
unexplored spectroscopically because of its large size.

The richness of kinematical signatures found in these studies motivated
us to search for similar evidence among the newly released Gaia DR1 data
(Gaia Collaboration 2016a,b), the so-called TGAS catalogue.
Since the Tycho catalogue on which
the Gaia DR1 is based only reaches down to $V \sim 12$, we do not
expect to find the majority of the individual stars in the kinematical
populations from Jeffries \e (2014) and Sacco \e (2015), since they
are mostly fainter than this ($11<V<19$). However, as we describe below, the Gaia
proper-motion data show clear evidence of a double population in Vela OB2,
with intriguing analogies with those from the Gaia-ESO Survey data.

\begin{figure}
\resizebox{\hsize}{!}{
\includegraphics[bb=5 10 485 475]{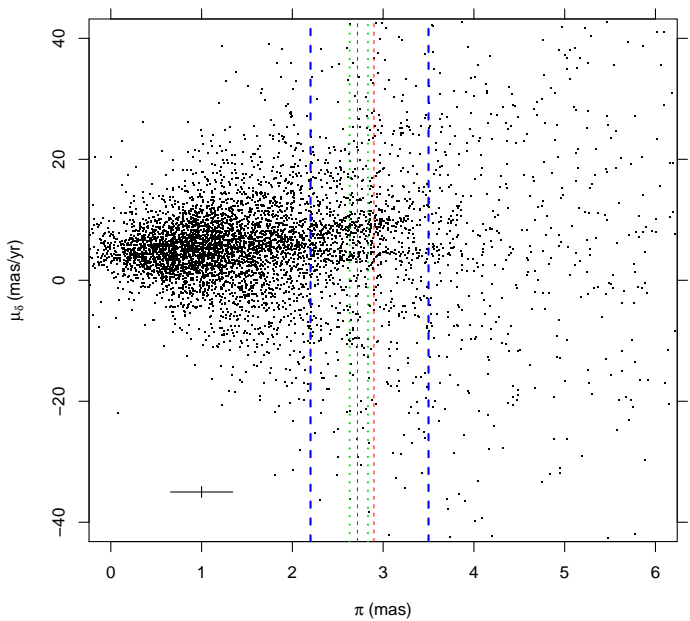}}
\caption{Proper motion along Declination $\mu_{\delta}$ vs.\ parallax $\pi$,
for the entire Vela OB2 TGAS sample.
Vertical lines correspond to literature distances for $\gamma^2$~Vel (red)
and NGC~2547 (green), and to our sample selection (blue).
The average errors for the parallax-selected sample are shown.
\label{plx-pmdec}}
\end{figure}

\begin{figure}
\resizebox{\hsize}{!}{
\includegraphics[bb=5 10 485 475]{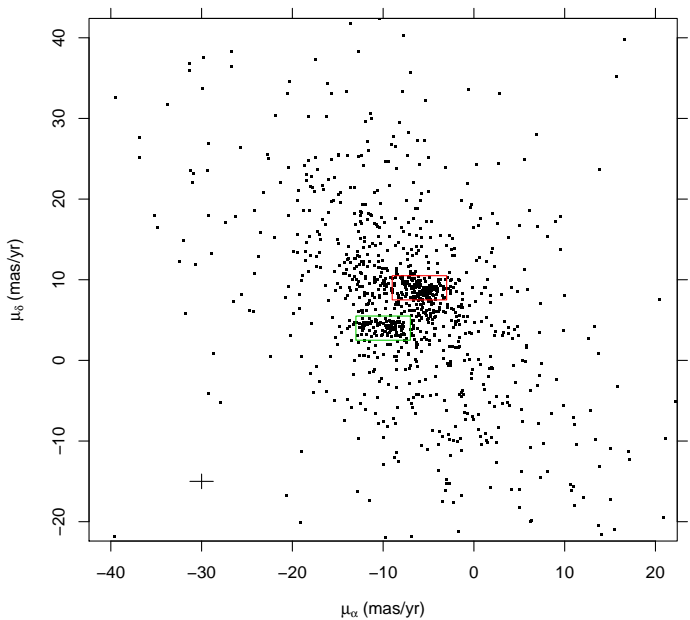}}
\caption{Proper-motion diagram for the parallax-selected subsample.
The red and green rectangles are used to select the two kinematical
populations C and~D, respectively.
The average errors for the parallax-selected sample are shown.
\label{pmra-pmdec}}
\end{figure}

\section{The Gaia data}
\label{data}

From the TGAS catalogue, we retrieved all entries within a radius of
4 degrees from the midpoint between \gv\ and NGC~2547 center, i.e.\ a
center of coordinates $(\alpha,\delta)=(122.4305,-48.27079)$. This sample
comprises of 5785 TGAS objects.
A diagram
showing proper motion along Declination (\pmdec) vs.\ parallax ($\pi$)
is shown in Figure~\ref{plx-pmdec}.  Close to the literature distances of \gv\
and NGC~2547 a characteristic pattern is found, with stars tending to
regroup in two horizontal, narrow bands in the diagram. Therefore, we
selected the subset of data within parallax limits $2.2 < \pi < 3.5$ mas
(1084 stars),
as suggested by the horizontal spread of the bands in the Figure. These
rather generous limits are justified since the TGAS precision on parallax
is much lower than that anticipated for the final Gaia data releases.
{
The mean uncertainty on $\pi$ for the parallax-selected sample is 0.34
mas. However, the TGAS parallaxes are likely to have an additional
systematic uncertainty of 0.3 mas (Gaia collaboration 2016b; Lindegren
et al.\ 2016), and if we add this in quadrature our chosen parallax
interval corresponds to a $\pm 1.4\sigma$ range, that would contain
about 85 per cent of stars that are actually at the distance of
$\gamma^2$ Vel and NGC~2547, if the uncertainties are normally
distributed.
}

A proper-motion diagram (\pmra, \pmdec) for this parallax-selected subset
is shown in Figure~\ref{pmra-pmdec}. This shows clearly the existence
of two localized overdensities of datapoints against a much broader
distribution of points (field stars). We therefore define from this diagram
(and from the previous parallax limits) two populations, termed here
C ($-9 < \mu_{\alpha} < -3$ mas/yr and $7.5 < \mu_{\delta} < 10.5$ mas/yr),
and~D ($-13 < \mu_{\alpha} < -7$ mas/yr and $2.5 < \mu_{\delta} < 5.5$ mas/yr).
We use C and D to avoid confusion with the two populations (A and B)
defined by Jeffries \e (2014); we discuss below the possible connection
between these pairs.
Population~C (D) comprises 148 (101) stars.
{
We have tried to estimate the field-star contamination in these samples
by considering TGAS stars, within the same parallax limits, falling in a
rectangular region of the (\pmra, \pmdec) diagram intermediate between
those enclosing populations~C and~D, that is $-13 < \mu_{\alpha} < -3$
mas/yr and $5.6 < \mu_{\delta} < 7.4$ mas/yr: 53 stars are found.
This region has the same area in the (\pmra, \pmdec) plane as those
enclosing populations~C and~D, so no area corrections are required.
Therefore, the expected level of contamination is $\sim 36$\% for
population~C, and $\sim 52$\% for population~D.
}

\begin{figure}
\resizebox{\hsize}{!}{
\includegraphics[bb=5 10 485 475]{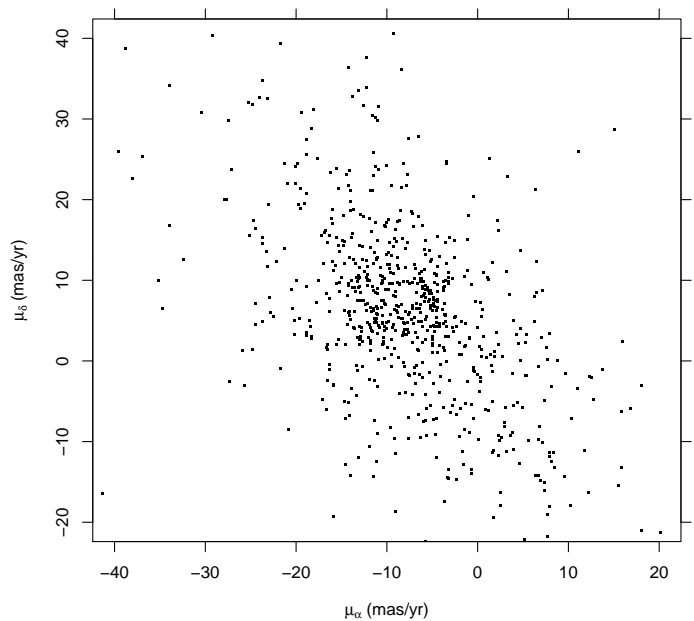}}
\caption{As in Figure~\ref{pmra-pmdec}, but using data from the external regions
in Vela OB2 (772 stars).
\label{pmra-pmdec-offax}}
\end{figure}

To help understand the spatial region where Populations~C and~D exist,
we have repeated the same selection of TGAS stars, but now
from an annulus surrounding the previous region, of $4^{\circ}$
($6^{\circ}$) inner (outer) radius,
which yielded the proper-motion diagram (\pmra, \pmdec) shown in
Figure~\ref{pmra-pmdec-offax}. Here, the two density peaks are no longer
clearly recognizable: the two populations~C and~D are therefore mostly confined
within the $4^{\circ}$-radius circle (nearly 50~pc in diameter at the
distance of \gv).

\begin{figure}
\resizebox{\hsize}{!}{
\includegraphics[bb=5 10 485 475]{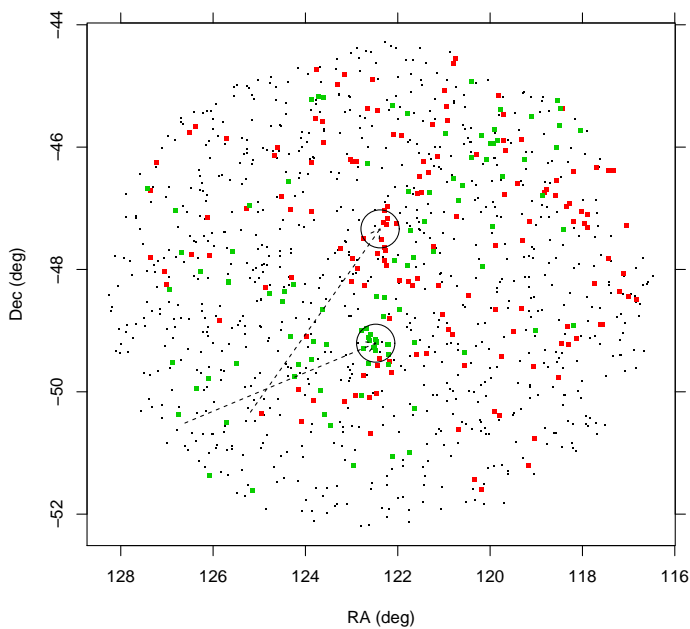}}
\caption{Spatial distributions of kinematical populations C and~D (red
and green respectively). The small black dots indicate the datapoints outside
of rectangles in Figure~\ref{pmra-pmdec}.
The big circles indicate positions of $\gamma^2$~Vel (north) and NGC~2547
(south).
The dashed arrows indicate their inferred positions 1.1~Myr ago.
\label{spatial}}
\end{figure}

Inside this radius, however, the selected stars from populations C and D are
distributed very inhomogeneously, as shown in Figure~\ref{spatial}:
{
the strongest clustering is found for a subset of
population D stars (green dots), spatially coincident
}
with the location of the NGC~2547 cluster (southern black circle
in Fig.~\ref{spatial}).  This fact suggests that our population D is
physically the same as the main kinematical population of this cluster,
found by Sacco \e (2015).  Population C is instead not particularly
clustered in or around the field studied by Jeffries \e (2014);
{
however, its proper motion is very similar to that of both populations~A
and~B, from UCAC4 data (Jeffries \e 2014), which have themselves
indistinguishable proper motions within the UCAC4 errors
(mean (\pmra,\pmdec) equal to (-5.9 $\pm$ 0.8, +8.5 $\pm$ 1.0) and
(-4.6 $\pm$ 1.0, +8.7 $\pm$ 0.9) mas/yr for populations~A and~B, respectively).
We therefore tentatively associate population~C with the \gv\ populations~A
and~B cumulatively, and population~D with NGC~2547.
}
Figure~\ref{spatial} also
shows that NGC~2547 is not the geometrical centroid of either population
C or D, which are spread more towards the northern half of the surveyed
region than south of NGC~2547. Such non-spherical geometry might suggest
that radial expansion from a small region was not the origin of the
large spatial extent of these populations.

\begin{figure}
\resizebox{\hsize}{!}{
\includegraphics[bb=5 10 485 475]{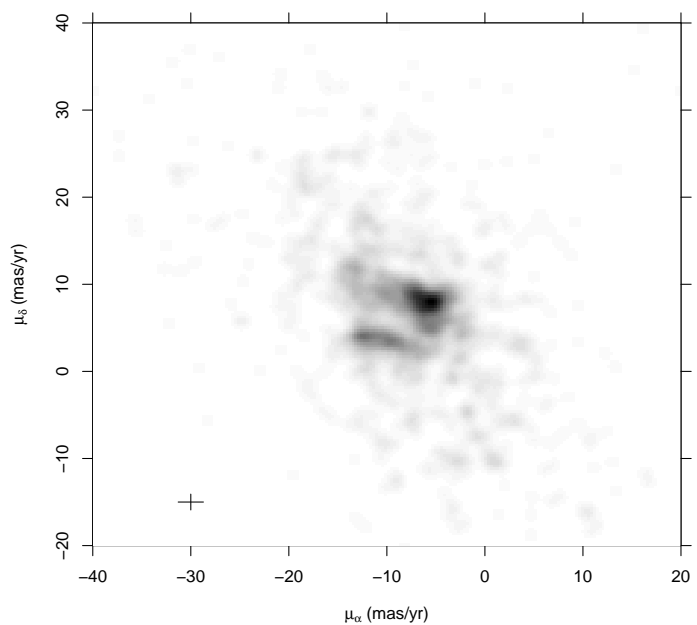}}
\caption{Smoothed spatial distributions of all datapoints except for
concentrated sub-populations C and~D.
\label{pmra-pmdec-sparse-smooth}}
\end{figure}

{
There is only a very small number of common objects between populations
C and~D and stars in Jeffries \e (2014: two matches), or Sacco \e (2015:
no matches), because of the small overlap between the TGAS and Gaia-ESO
magnitude ranges.
There are no additional matches even considering the Gaia-ESO targets in
the NGC~2547 field not published in Sacco \e (2015).
}
The two matches with Jeffries \e (2014) are stars
TYC~8140-2731-1 and TYC~8140-6234-1 (2MASS J08094701-4744297 and
J08092627-4731001, respectively), both belonging to population~C.
Jeffries \e assign the second star to their population~B (5\%
probability of belonging to population~A), while the population of the
first one is undetermined.

{
The contrast between the centrally-condensed distributions of members of
\gv\ and NGC~2547 clusters, and the sparseness of both populations~C
and~D renders questionable a one-to-one correspondence between the former
and the latter. One possibility is then that each of populations~C and~D
is actually composed of a concentrated and a more sparse component.
To this aim we
have considered separately, for each of populations C and~D, stars
within and outside a $1^{\circ}$-radius circle around their respective
centers (assumed to be the centers of \gv\ and NGC~2547 clusters).
In doing so, the number statistics is reduced, and the relative level of
field-star contamination becomes more important.
The just-defined concentrated (sparse) population~C contains 24 (124) stars,
which likely contain 4 (49) contaminants.
For population~D we obtain 27 (74) concentrated (sparse) stars,
of which 2 (51) are likely to be contaminants.
It can therefore be seen that, although the sparse sub-populations
contain in both cases the majority of members, their contamination level
is far above those of the concentrated sub-populations.
The number of field contaminants for each of the sparse sub-populations
can be modeled as a random distribution with mean $\mu \sim 50$ and
std.dev.\ $\sigma \sim 7$ stars. The contamination-corrected numbers of stars
in sparse sub-populations~C and~D (74 and 24 stars, respectively) would
be significant at the 10.6$\sigma$ and 3.4$\sigma$ level, respectively.
{
Although the existence of the sparse population~D is
only slightly above a safe confidence threshold,
the distribution of datapoints in the (\pmra, \pmdec)
plane after subtraction of the concentrated C and D sub-populations
(Figure~\ref{pmra-pmdec-sparse-smooth}),
still provides evidence for both C and D sparse sub-populations.
Therefore, even if the properties of
the sparse population~D are ill-defined because of strong contamination,
its existence is not greatly in doubt.
}

It may also be of some importance to consider the subsets of TGAS stars
in our populations with Hipparcos data, and characterized by proper-motion
measurements that are more precise by an order of magnitude than the
other TGAS stars. These are 1 star (out of 24) for concentrated population~C,
9 stars (out of 124) for sparse population~C,
6 (out of 27) for concentrated population~D, and finally
5 (out of 74) for sparse population~D.
The sparse population~D is thus characterized by both a low
number and small percentage of higher-quality proper-motion measurements,
which adds to the difficulties in assessing its
properties. For example, a comparison between the proper-motion
distributions of sparse vs.\ concentrated D~sub-populations would be
vitiated by their different error distributions.

}

{
We have examined if any indications of expansion or contraction can be
found from the TGAS data. Rather surprisingly, a contraction pattern for
population~C (only) is
seen in $RA$ (\pmra\ negatively correlated with $\Delta(RA)$ from
\gv\ center), but not in $Dec$, significantly above the
quoted proper-motion random errors. However, the TGAS data in the
region are affected by a known systematic correlation between exactly
the same parameters, which might be entirely responsible for the
observed effect.
}

\begin{figure}
\resizebox{\hsize}{!}{
\includegraphics[bb=5 10 485 475]{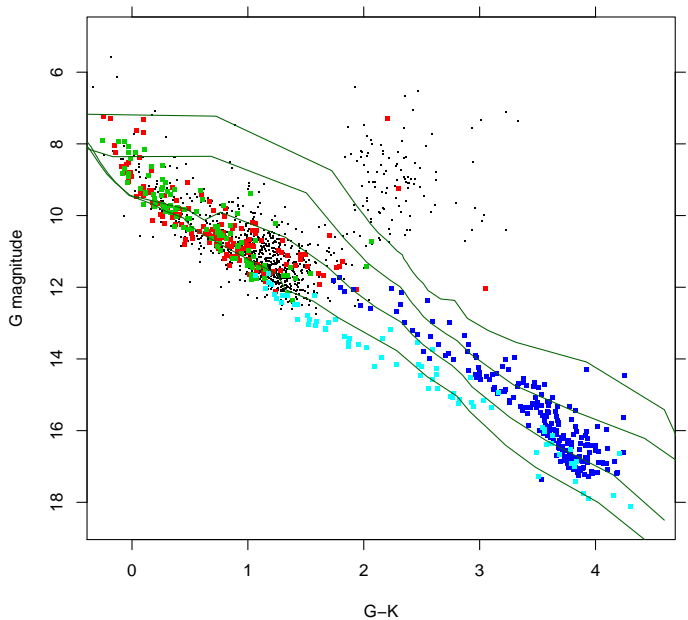}}
\caption{$(G,G-K)$ color-magnitude diagram.
Symbols are the same as in Fig.~\ref{spatial}, with the addition of
stars from Jeffries \e (2014 - blue) and Sacco \e (2015 - cyan).
Shown are isochrones from Siess \e (2000) (dark green), at ages
of 1, 3, 10, and 30~Myr, at zero reddening and distance of 356~pc.
\label{g-gk}}
\end{figure}

\section{Discussion}
\label{discuss}

\subsection{Age of kinematical populations}
\label{age}

{
We have studied the placement of populations C and~D on a color-magnitude
diagram (CMD). We use the Gaia $G$ and the 2MASS $K_s$ band magnitudes
to build the CMD shown in Figure~\ref{g-gk}.
In the same Figure we also show the \gv\ and NGC~2547 cluster members
from Jeffries \e (2014) and Sacco \e (2015), respectively.
}

Very few of our population-C or -D stars may be giants;
on the other hand, these populations are rich in upper main-sequence stars.
These features
are consistent with a young age, like that of
\gv\ and NGC~2547 clusters.
A match with the SIMBAD database finds several tens matches for both
population-C and -D stars, mostly of spectral type A or B.
Therefore, we have overlaid the
CMD in Figure~\ref{g-gk} with pre-main-sequence evolutionary tracks from
Siess \e (2000), converted to the Gaia $G$ band following Jordi \e (2010).
These isochrones suggest that
the ages of populations C and D are consistent with those of the \gv\
and NGC~2547 clusters. Some indication is also found of a slightly older
age for population-D stars (again, best consistent with the NGC~2547 age)
with respect to population-C stars (more similar to \gv\ cluster stars).
{
To check this, we have performed a Kolmogorov-Smirnov test comparing the
respective magnitude differences $\Delta G$ above the 30~Myr isochrone,
in the $G-K$ range [1.0-2.0] which is most sensitive to ages in the
10-30~Myr range. The probability that the two samples come from the same
parent population is found to be $p=0.7$\%. The same test applied to the
$G-K$ range [1.0-1.7] gives $p=4.9$\%. Contamination of course acts in
the sense of diluting actual differences, so we may conclude that there
is a real age difference between populations C and D.
}

\subsection{The recent past of populations C and D}
\label{formation}

{
Using their well-defined mean proper motions, we may trace back in time
the apparent average positions of populations C and~D over a few Myrs.
}
The arrows in Figure~\ref{spatial} show the
positions of \gv\ and NGC~2547 extrapolated back 1.1~Myr, using the
{
(error-weighted) mean proper motions of concentrated populations C and~D,
having values (\pmra, \pmdec) = (-6.12, 9.80) and (-9.00, 4.27)~mas/yr,
respectively.
}
This shows that their (sky-projected) distance is now slowly increasing,
{
while it reached a minimum value of 0.65 degrees (4.04~pc) 1.1~Myr ago.
This is a small enough value, compared to the apparent sizes of both
\gv\ and NGC~2547 clusters, to expect a significant dynamical
interaction between them, provided that also their distances from the Sun
are (or were at that time) coincident within a few pc.
}
{
An intriguing possibility is therefore that population~B in the \gv\
cluster was originated from a tidal stripping event during a close encounter
with the (denser) cluster NGC~2547. A difficulty with this hypothesis
lies in the offset of the RV distribution of population~B with respect to
that of population~A, whereas tidal stripping may be expected to create
two symmetric (leading and trailing) tails.
}

{
To summarize, the Gaia DR1 data indicate that populations C and~D had
the highest probability of a mutual interaction in the past 1-1.5~Myr.
}
{
Determining if this has actually taken place depends on a more accurate
determination of the populations' respective parallaxes, which are
expected from future Gaia data releases.
}

{
{
The existence of the sparse population~D appears sufficiently certain
from the TGAS data; however, its properties could not be studied in any
detail, because of the small number statistics and
large field-star contamination.
}
{
Future Gaia data releases
will permit to determine its properties more accurately.
}

\begin{acknowledgements}
We acknowledge useful suggestions from an anonymous referee.
This work has made use of data from the European Space Agency (ESA)
mission {\it Gaia} (\url{http://www.cosmos.esa.int/gaia}), processed by
the {\it Gaia} Data Processing and Analysis Consortium (DPAC,
\url{http://www.cosmos.esa.int/web/gaia/dpac/consortium}). Funding
for the DPAC has been provided by national institutions, in particular
the institutions participating in the {\it Gaia} Multilateral Agreement.
This research has made use of the VizieR catalogue access tool,
and of the cross-match service provided by CDS, Strasbourg, France. 
\end{acknowledgements}

\bibliographystyle{aa}

\end{document}